\documentstyle[epsfig,12pt]{article}
\begin{document}

\title{{\bf Reconstructing an economic space from a market metric}}
\author{{\bf R. Vilela Mendes} \and {\small Grupo de F\'{i}sica Matem\'{a}tica, Av.
Gama Pinto 2,} \and {\small \ 1699 Lisboa Codex, Portugal
(vilela@cii.fc.ul.pt)} \and {\bf Tanya Ara\'{u}jo and Francisco Lou\c{c}\~{a}%
} \and {\small Departamento de Economia, ISEG,} \and {\small R. Miguel Lupi
20, 1200 Lisboa, Portugal} \and {\small (tanya@iseg.utl.pt,
flouc@iseg.utl.pt)}}
\date{}
\maketitle

\begin{abstract}
Using a metric related to the returns correlation, a method is proposed to
reconstruct an economic space from the market data. A reduced subspace,
associated to the systematic structure of the market, is identified and its
dimension related to the number of terms in factor models. Example were
worked out involving sets of companies from the DJIA and S\&P500
indexes.

Having a metric defined in the space of companies, network topology
coefficients may be used to extract further information from the data. A
notion of ``continuous clustering'' is defined and empirically related to
the occurrence of market shocks.
\end{abstract}

{\bf Keywords}: Market metric, Economic space, Factors

\section{Introduction}

In spite of the important achievements obtained in finance theory (see for
example http://welch.som.yale.edu/academics/toptenfinance.html and ch.35 in
Ref.\cite{Brealey}) nobody claims that the fundamental laws of the economic
process are known. A set of fundamental laws under which all economic
relations might be interpreted is certainly not known and, even if such laws
were to exist, we do not know how to infer from the data what are the variables
that play the relevant role in the equations. Instead, economic theory generally
establishes, a priori, the models as sets of restrictions in order to
proceed to statistical tests of the data. Most of the developments in
finance theory follow this line.

The dominant views, such as the Efficient Market Hypothesis, based on the
work of Samuelson\cite{Samuelson} and Fama\cite{Fama2}, and the derived
models, such as the multifactor capital asset pricing model (CAPM) \cite
{Merton} and arbitrage pricing theory (APT) \cite{Ross1}, assess the
evolution of financial markets as the result of the rational action of informed
agents faced with Brownian processes. These models provide conceptual
insights on the issues of pricing and portfolio selection, although attempts
to test them has been hindered by the inability to find a reliable set of
factors to explain the securities return data. Chen et al. \cite{Ross2} have
attempted to establish statistical correlations between some economic facts
(like unanticipated changes in industrial production, interest rates or
inflation) and asset returns, to identify the {\it economic forces} that are
driving the market. But the very identification of such forces, and the
rationale for its theoretical underpinnings, is also controversial.

Mandelbrot, who studied the properties of stable distributions other than
the Gaussian, applied new statistical methods to financial series,
suggested the existence of low frequency dependence in the stock market data
and challenged the dominance of Brownian processes\cite{Mandelbrot}. Indeed,
Mandelbrot interpreted the fat tails in the distribution of changes of
prices and the empirical evidence of sharp discontinuities in the evolution
of these markets as evidence for the presence of a stable distribution.
Instead, his critics argued that the financial series should be interpreted
as a result of variables with typically high frequency variance, such as
serial correlation and Markov dependence. Consequently the fat tails of the
distribution of price changes could be explained by subordinate stochastic
processes, in particular by time varying variances of Gaussian processes,
rather than by stable distributions or truncated L\'{e}vy processes.

Mandelbrot's views were understood as a criticism of the conventional wisdom
on the inexistence of structure in the evolution of stock markets, and were
generally rejected. Also based on the market, we will address here this topic 
of debate from a different approach. Instead of establishing
correlations with predefined factors, our point of view is that it may be
possible to extract from the data itself, if not the economic variables, at
least their geometrical relations. And also that such an exploration might
be fruitful for statistical analysis. The idea is simply stated in the
following terms:

(i) Pick a representative set of $N$ stocks and their historical data of
returns over some time period.

(ii) From the returns data, using an appropriate metric, compute the matrix
of distances between the $N$ stocks.

The problem now is reduced to an embedding problem where, given a set of
distances between points, one asks what is the smallest manifold that
contains the set. Given a graph G and an allowed distortion there are
algorithmic techniques\cite{Linial} to map the graph vertices to a normed
space in such a way that distances between the vertices of G match the
distances between their geometric images, up to the allowed distortion.
However, these techniques are not directly applicable to our problem
because in the distances between assets, computed from their return
fluctuations, there are systematic and unsystematic contributions.
Therefore, to extract factor information from the market, we have somehow to
separate these two effects. The following stochastic geometry technique is
used:

(iii) From the matrix of distances, compute coordinates for the stocks in an
Euclidean space of dimension $N-1$. (For a degenerate matrix the embedding
dimension may be smaller)

(iv) The stocks are now represented by a set $\left\{ x_{i}\right\} $ of
points in $R^{N-1}$, to which we assign masses $\left\{ m_{i}\right\} $
equal to their market capitalizations.

(v) To this cloud of weighted points we apply the standard analysis of
reduction of their coordinates to the center of mass and computation of the
eigenvectors of the inertial tensor.

(vi) The same technique is also applied to surrogate data, namely to data
obtained by independent time permutation for each stock and to random data
with the same mean and covariance.

(vii) The eigenvalues in (v) are compared with those of (vi). The directions
for which the eigenvalues are significantly different are now identified as
the market systematic variables.

Using weights (masses) proportional to the market capitalizations we are
attempting to identify{\it \ the empirically constructed variables} that
drive the market and the number of surviving eigenvalues is the {\it %
effective dimension} of this economic space. Of course, what such a
procedure reconstructs is the economic space associated to the set of stocks
that is considered, not to the full market. Even if a very large set of
financial assets is used, there is no implied claim that financial markets
fully reflect all what we would like to know about macroeconomics. All one
is trying to do here is to reconstruct {\it an} economic space, not {\it the}
economic space.

The same technique may be used to infer factors for portfolio hedging
purposes. In this case there is no reason to include weights and all
companies may be considered to have the same weight. We will have examples
of both types of calculation.

In a recent paper Gopikrishnan et al.\cite{StanleyBS} used similar
techniques, although with a different perspective. Diagonalizing the
correlation matrix (which is related to the metric we use) they have tried
to identify particular eigenvectors with the traditional industrial sectors.
In our analysis the economic dimensions may or may not correspond to
economic sectors or to other known economic factors or to any combination of 
them. It is up to the data to say what they are, independently of any previously
established concepts.

In Section 2 the method is explained in detail and then, as an example, it
is applied to market data of a set of large companies that are or have been
in the Dow Jones Industrial Average and S\&P500 indexes.

\section{Reconstruction of an economic space}

\subsection{The market metric}

From the returns $r(k)$ for each security 
\begin{equation}
r_{t}(k)=\log (p_{t}(k))-\log (p_{t-1}(k))  \label{2.1}
\end{equation}
one defines a normalized vector 
\begin{equation}
\overrightarrow{\rho }(k)=\frac{\overrightarrow{r}(k)-\left\langle 
\overrightarrow{r}(k)\right\rangle }{\sqrt{n\left( \left\langle 
\overrightarrow{r}^{2}(k)\right\rangle -\left\langle \overrightarrow{r}%
(k)\right\rangle ^{2}\right) }}  \label{2.2}
\end{equation}
$n$ being the number of components (number of time labels) in the vectors $%
\overrightarrow{r}(k)$. With this vector one defines the {\it distance}
between the securities $k$ and $l$ by the Euclidean distance of the
normalized vectors, 
\begin{equation}
d_{kl}=\left\| \overrightarrow{\rho }(k)-\overrightarrow{\rho }(l)\right\| =%
\sqrt{2\left( 1-C_{kl}\right) }  \label{2.3}
\end{equation}
$C_{kl}$ being the correlation coefficient of the returns 
\begin{equation}
C_{kl}=\frac{\left\langle \overrightarrow{r}(k)\overrightarrow{r}%
(l)\right\rangle -\left\langle \overrightarrow{r}(k)\right\rangle
\left\langle \overrightarrow{r}(l)\right\rangle }{\sqrt{\left( \left\langle 
\overrightarrow{r}^{2}(k)\right\rangle -\left\langle \overrightarrow{r}%
(k)\right\rangle ^{2}\right) \left( \left\langle \overrightarrow{r}%
^{2}(l)\right\rangle -\left\langle \overrightarrow{r}(l)\right\rangle
^{2}\right) }}  \label{2.4}
\end{equation}
Being an Euclidean distance between two vectors, Eq.(\ref{2.3}) satisfies
the usual distance axioms. It is the distance between market securities that
was proposed in \cite{Mantegna1} and \cite{StanleyM}.

This distance is related to the covariances and much of what we discuss
below could be carried out in a purely statistical setting. However the fact
that $d_{kl}$ is a properly defined distance gives a meaning to geometric
notions and geometric tools in the study of the market.

\subsection{Characteristic dimensions, systematic covariance and factors}

After the distances are computed, for the set of $N$ securities, they are
imbedded in $R^{N-1}$ with coordinates $\left\{ \overrightarrow{x}%
(k)\right\} $. The center of mass $\overrightarrow{R}$ is computed, 
\begin{equation}
\overrightarrow{R}=\frac{\sum_{k}m_{k}\overrightarrow{x}(k)}{\sum_{k}m_{k}}
\label{2.5}
\end{equation}
the coordinates reduced to the center of mass, 
\begin{equation}
\overrightarrow{y}(k)=\overrightarrow{x}(k)-\overrightarrow{R}  \label{2.6}
\end{equation}
and the inertial tensor 
\begin{equation}
T_{ij}=\sum_{k}m_{k}y_{i}(k)y_{j}(k)  \label{2.7}
\end{equation}
is diagonalized, the set of eigenvalues and normalized eigenvectors being $%
\left\{ \lambda _{i},\overrightarrow{e_{i}}\right\} $. The eigenvectors $%
\overrightarrow{e_{i}}$ define the characteristic directions of the weighed
set of securities and their $z_{i}(k)$ coordinates along these directions
are obtained by projection 
\begin{equation}
z_{i}(k)=\overrightarrow{y}(k)\bullet \overrightarrow{e_{i}}  \label{2.8}
\end{equation}

As stated before, the most relevant characteristic directions for our
purposes are those that correspond to the eigenvalues which are clearly
different from those obtained from surrogate or random data. They define a
subspace $V_{d}$ of dimension $d$. This $d-$dimensional subspace carries the
(systematic) information related to the market correlation structure.

In portfolio optimization models of the mean-variance type, one usually
distinguishes between the systematic and unsystematic (or specific)
contributions to the portfolio risk. The former are associated to the
correlations between the assets in the portfolio and the latter to the
individual variances alone. Using our construction we find that
part of the correlations contribution is indistinguishable from random
data. Hence the market (systematic) structure is carried by a smaller $d-$%
dimensional subspace. This suggests the definition of a {\it market
dimension }$d$ and a {\it systematic covariance.}

Denote by $\overrightarrow{z}(k)^{(d)}$ the restriction of the $k-$asset to
the subspace $V_{d}$. and by $d_{kl}^{(d)}$ the distances restricted to this
space. Then using Eqs.(\ref{2.3}) and (\ref{2.4}) we may define a notion of 
{\it systematic covariance }$\sigma _{kl}^{(d)}$%
\begin{equation}
\sigma _{kl}^{(d)}=\mu _{k}\sqrt{\sigma _{kk}-\overline{r}_{k}^{2}}\mu _{l}%
\sqrt{\sigma _{ll}-\overline{r}_{l}^{2}}\left( 1-\frac{1}{2}\left(
d_{kl}^{(d)}\right) ^{2}\right)  \label{2.9}
\end{equation}
where $\mu _{k}=|\overrightarrow{z}(k)^{(d)}|/|\overrightarrow{z}(k)|$ , $%
\overline{r}_{k}=\left\langle \overrightarrow{r}(k)\right\rangle $ and $%
\sigma _{kk}=\left\langle \overrightarrow{r}(k)\overrightarrow{r}%
(k)\right\rangle $ .

In a portfolio optimization problem 
\begin{equation}
r=\sum_{k}W_{k}r(k)  \label{2.10}
\end{equation}
the function to be minimized would be 
\begin{equation}
\sum_{k\neq l}\sigma _{kl}^{(d)}W_{k}W_{l}+\sum_{k}\sigma _{kk}W_{k}^{2}
\label{2.11}
\end{equation}
identical to the classical Markowitz problem, but with the systematic
covariance part restricted to the subspace $V_{d}$.

This analysis also provides a rationale for the choice of the number of
terms in the construction of factor models, the factors being constructed
from the leading characteristic dimensions (see the example below).

\subsection{Clustering}

In addition to a detailed subspace analysis of the economic space, existence
of a market metric provides network topology coefficients to characterize
the whole space. One such notion is {\it clustering}, a meaningful
well-known notion in graph theory. Using the distance matrix $d_{ij}$ (Eq.(%
\ref{2.3})) to construct the minimal spanning tree connecting the $N$
securities, as in Mantegna \cite{Mantegna1}, we might then apply the graph
theoretical notion of clustering to the spanning tree. However this
construction neglects part of the information contained in the distance
matrix. Instead we introduce a notion of {\it continuous clustering} as
follows:

$d_{ij}$ being the distance between the securities $i$ and $j$ and $%
\overline{d}$ the average distance we define a function 
\begin{equation}
V_{ij}=\exp \left( -\frac{d_{ij}}{\overline{d}}\right)  \label{2.12}
\end{equation}
which represents the {\it neighbor degree} of the securities $i$ and $j$. A
(continuous) clustering coefficient is then defined by 
\begin{equation}
C=\frac{1}{N(N-1)(N-2)}\sum_{i\neq j\neq k}V_{ij}V_{jk}V_{ik}  \label{2.13}
\end{equation}

\section{An example}

We have considered the following 34 large companies which are, or have been,
in the Dow Jones Industrial Average (DJIA) index:

Alcoa (AA), Honeywell (HON), American Express (AXP), AT\&T (T), Boeing (BA),
Caterpillar (CAT), Chevron (CHV), Coca-Cola (KO), Dupont Nemours (DD),
Eastman Kodak (EK), Exxon (XON), General Electric (GE), Goodyear (GT), IBM
(IBM), International Paper (IP), McDonalds (MCD), Merck (MRK), Minnesota
Mining (MMM), General Motors (GM),Philip Morris (MO), Procter \& Gamble
(PG), Sears (S), Texaco (TX), United Technologies (UTX), Citigroup (C),
Hewlett-Packard (HWP), Home Depot (HD), Intel (INTC), J. P. Morgan Chase
(JPM), Johnson \& Johnson (JNJ), Microsoft (MSFT), SBC Communications (SBC),
Wal-Mart (WMT), Walt Disney (DIS).

They will be denoted by their tick symbols and we use daily data for
the time period from September 1990 to August 2000.

\begin{figure}[htb]
\begin{center}
\psfig{figure=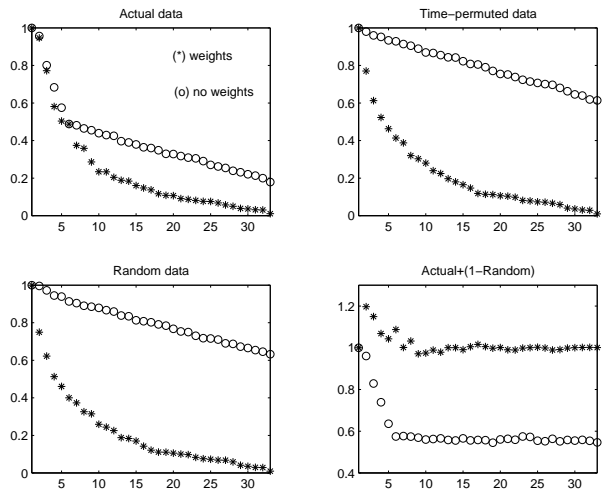,width=8truecm}
\end{center}
\caption{Eigenvalue distributions for the actual, 
time-permuted and random data}
\end{figure}

Using the whole data for the ten years, to define the vectors $%
\overrightarrow{\rho }(k)$ for each company, the calculations described in
Section 2 have been performed for the actual returns data, for the
time-permuted data and for random data with the same mean and variance as
the actual data. In all cases we have performed the calculations with and
without weights. The ordered eigenvalue distributions that were obtained are
shown in Fig.1. The conclusion is that the (systematic) market structure is
contained in the first five dimensions. That is, these dimensions capture
the structure of the deterministic correlations and economic trends that are
driving the market, whereas the remainder of the market space may be
considered as being generated by random fluctuations. For this market, these
five dimensions define our {\it empirically constructed economic variables}.

\begin{figure}[htb]
\begin{center}
\psfig{figure=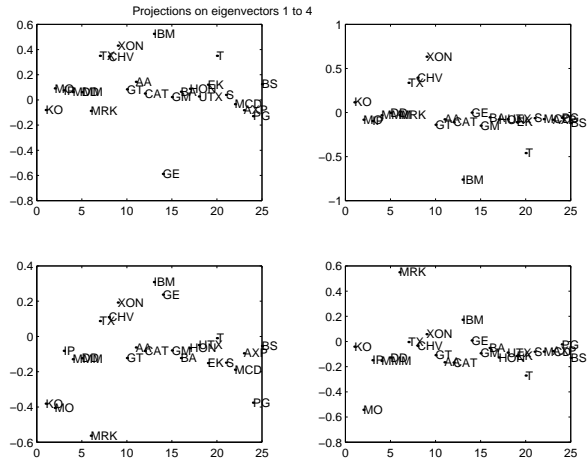,width=8truecm}
\end{center}
\caption{Projection of the
(weighted) stocks along the first four eigenvectors}
\end{figure}

\begin{figure}[htb]
\begin{center}
\psfig{figure=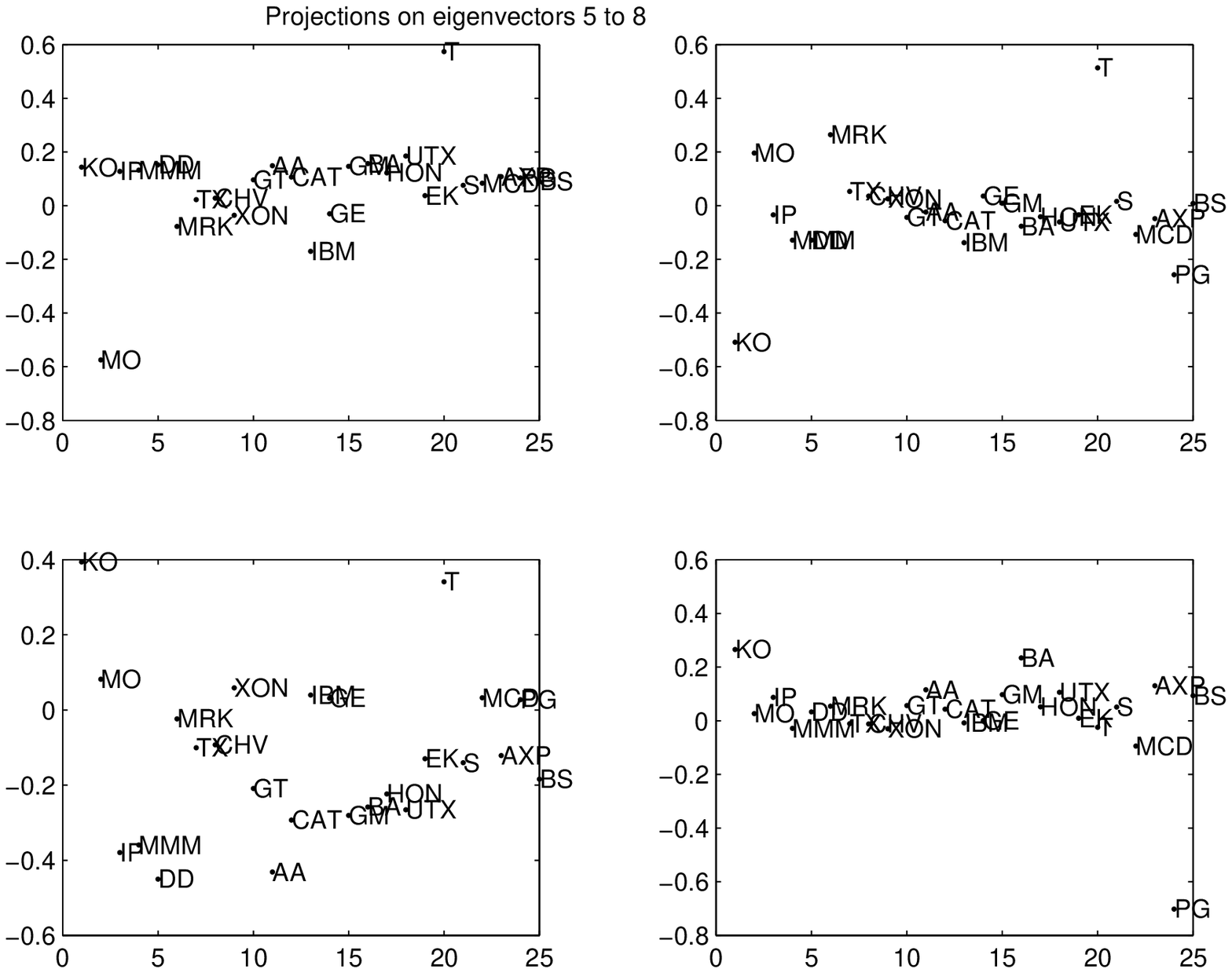,width=8truecm}
\end{center}
\caption{Projection of the
(weighted) stocks along the eigenvectors 5 to 8}
\end{figure}

To have a qualitative idea concerning the structure of the characteristic
dimensions, we have plotted in Figs. 2 and 3 the projections of the
(weighted) stocks along the directions of the first eight eigenvectors. In
the x-axis the companies are ordered according to their standard industrial
code. Although some companies in the same sector (for example the oil
companies) have similar projections in the dominant eigenvalues, this is not
at all true for all sectors, nor all companies. The association of companies
working on different products in the same one or two-dimensional subspace is
a confirmation of the fact that the search for the factors that drive the
market cannot be identified with a definition of economic sectors. Notice
that to be in the same market subspace, does not mean to be close to each
other and some interesting anticorrelation effects are clear in Fig.2 and 3.
This may be important to develop portfolio hedging strategies.

\begin{figure}[htb]
\begin{center}
\psfig{figure=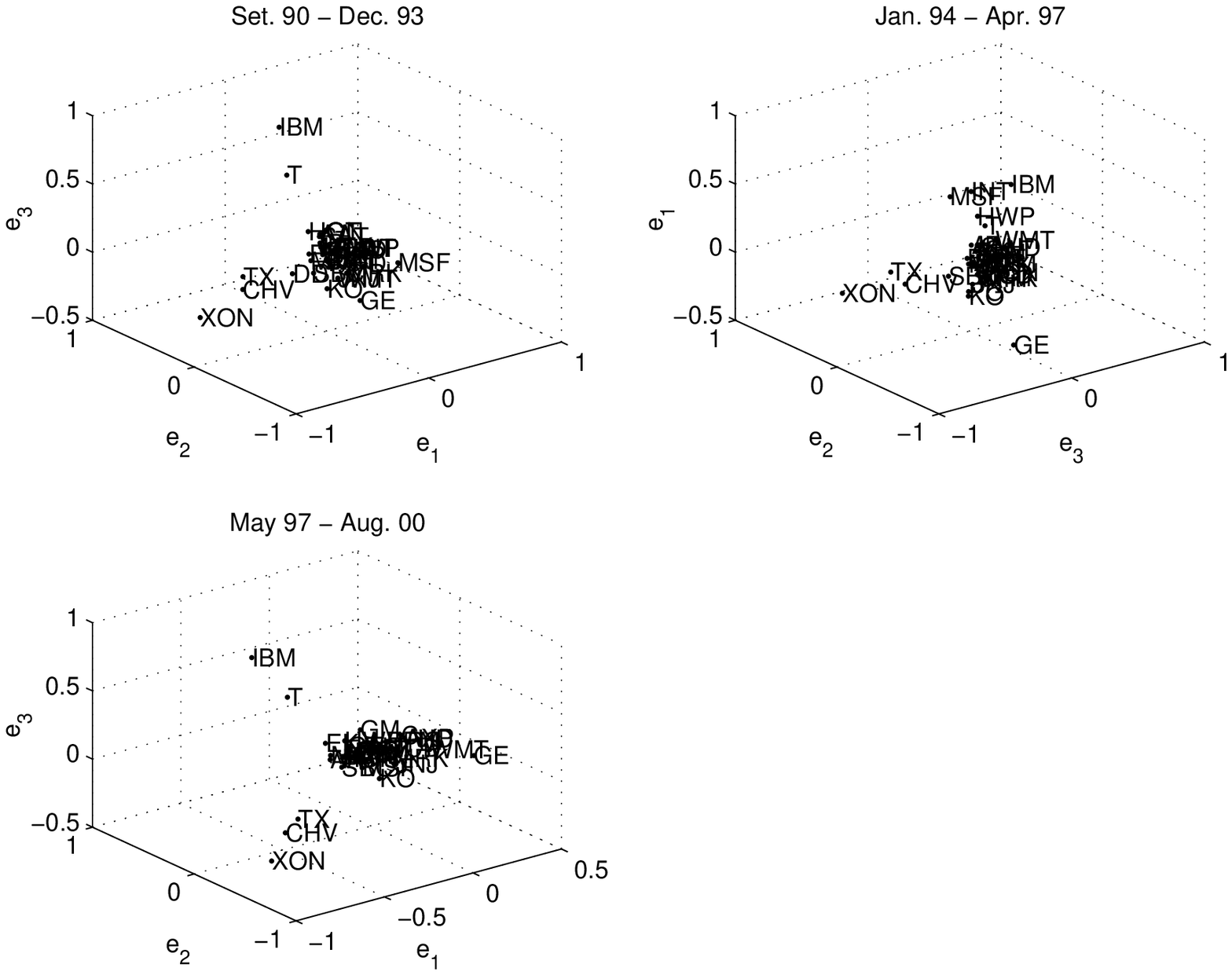,width=8truecm}
\end{center}
\caption{The leading 3-dimensional
subspaces associated to three chronologically successive periods}
\end{figure}

To test the stability of the economic structure inferred from the market, we
have divided the data in three chronologically successive batches and
performed the same operations. The behavior of the eigenvalue distributions
is very much the same. In Fig. 4 we have plotted the three dimensional
subspaces associated to the three largest eigenvalues. Apart from
statistical fluctuations, the reconstructed spaces show a reasonable degree
of stability. However, similarity of the figures is only apparent with a
permutation of the axis between the first and the second plot. The ordering
of the largest eigenvalues changes in time although the overall
distribution remains approximately the same. These ordering change may have
an economic meaning and be related to the relative importance and stability
of groups of companies in different periods of expansion or recession. What
is interesting, however, is the relative stability of the company positions
and the size and distribution of the eigenvalues. It is as if the effective
dimensionality of the space remained the same but with a pulsating effect on
its shape.

To test the dependence of the characteristic dimensions of the space on the
number of companies we have added to our set, data of the same ten years
period for 36 other large companies represented in the S\&P500 index,
namely (tick symbols only):

ABT, MHP, MEL, NYT, NKE, OXY, PEP, PFE, PHA, CBE, ADBE, APA, ASH, AAPL, BAC,
BK, BAX, BDK, CL, XRX, DCN, DAL, DG, SYY, F, G, HAL, EOG, HLT, RBK, SGP,
SLB, UNP, UIS, WHR, GDW.

Performing the same analysis as before for this larger set of 70 companies,
we have found that the number of relevant eigenvectors grows from five to
six. The small increase on the number of relevant characteristic dimensions
for a set with double the size of the first one, and which covers a wider
range of products, is quite remarkable. It seems to indicate that the
systematic factors in the market are relatively few and furthermore, that
they may be empirically defined.\bigskip

Finally we illustrate the computation of a set of empirical factors from the
geometrical analysis of the first set of 34 companies. A factor model for
the returns $r_{i}$ is 
\begin{equation}
r_{i}=a_{i}+\sum_{k=0}^{5}b_{ki}f_{k}+\varepsilon _{i}  \label{3.1}
\end{equation}
where the $a_{i}$ are called the intercepts, $b_{ki}$ the factor loadings
and $\varepsilon _{i}$ the residual random terms.

Recall that the first step in our analysis was the embedding of the 34
companies as a set of points in a 33-dimensional space. The company
coordinates are then reduced to the center of mass (Eq.(\ref{2.6})) and for
the computation of the factors we consider equal masses $m_{k}$. The vectors $%
y_{i}\left( t\right) $ denotes the time series reduced to the center of mass. 
\[
y_{i}\left( t\right) =r_{i}\left( t\right) -\overline{r}\left( t\right) 
\]
The zero-factor $f_{0}$ is simply the average 
\begin{equation}
f_{0}\left( t\right) =\overline{r}\left( t\right) =\frac{1}{34}%
\sum_{i=1}^{34}r_{i}\left( t\right)  \label{3.2}
\end{equation}
When the 5 relevant directions are identified, one obtains a 5-dimensional
subspace in a 33-dimensional space. The 5 factors are simply the 5
eigenvectors, associated to the largest eigenvectors, expressed in terms of
the time series of the companies. They are obtained as follows: Let $V$ be a
matrix with columns being the (center of mass) coordinates of the normalized
eigenvectors and $C$ a matrix containing as lines the (center of mass)
coordinates of companies 2 to 34. Then 
\[
M=CV 
\]
is a matrix containing, as lines, the (center of mass) company coordinates
projected on the eigenvectors. The factors, that is, the largest
eigenvectors written in terms of the time series of the companies are 
\[
f_{i}\left( t\right) =\sum_{n}M_{in}^{-1}y_{n}\left( 2:34\right) \left(
t\right) 
\]
where $y_{n}\left( 2:34\right) $ denotes the center of mass coordinates of
the companies 2 to 34.

Performing these operations on our data set, we have obtained vanishing $%
a_{i}$ intercepts ($\leq 10^{-7}$) and factor loadings $b_{ki}$ and
variances of the residual random terms $\varepsilon _{i}$ as listed below.
These variances are of order 50\% of the total variance of each company
return. This might be considered too high a value for a satisfactory factor
model. However it corresponds closely to the sum of the remaining 29
eigenvectors. These 29 eigenvalues are associated to dimensions which cannot
be distinguished from those of random data. Therefore one concludes that no
reliable improvement beyond the 5-factor model is possible with this data.

In Fig. 5 we have plotted the contribution ($M_{in}^{-1}$) of each time
series to the factors.

\newpage

\begin{center}
\begin{tabular}{|c|c|c|c|c|c|c|c|}
\hline
&  & b$_{1i}$ & b$_{2i}$ & b$_{3i}$ & b$_{4i}$ & b$_{5i}$ & $\sigma
^{2}(\varepsilon _{i})(10^{-4})$ \\ \hline
1 & AA & -0.240 & 0.266 & 0.229 & -0.171 & 0.162 & 2.15 \\ \hline
2 & HON & -0.159 & 0.143 & 0.149 & 0.011 & -0.014 & 2.57 \\ \hline
3 & AXP & 0.209 & -0.043 & 0.072 & -0.093 & -0.373 & 1.75 \\ \hline
4 & T & -0.094 & 0.501 & -0.335 & 0.551 & -0.269 & 1.29 \\ \hline
5 & BA & -0.045 & 0.029 & 0.030 & -0.042 & 0.008 & 3.02 \\ \hline
6 & CAT & -0.123 & 0.213 & 0.271 & -0.113 & 0.076 & 2.23 \\ \hline
7 & CHV & -0.609 & -0.348 & -0.241 & -0.082 & -0.067 & 0.96 \\ \hline
8 & KO & 0.129 & -0.342 & 0.102 & 0.182 & 0.098 & 2.00 \\ \hline
9 & DD & -0.198 & 0.051 & 0.322 & -0.080 & 0.127 & 2.02 \\ \hline
10 & EK & -0.032 & 0.137 & 0.004 & 0.124 & 0.299 & 2.76 \\ \hline
11 & XON & -0.547 & -0.408 & -0.218 & -0.054 & -0.045 & 1.01 \\ \hline
12 & GE & 0.217 & -0.169 & 0.089 & -0.108 & -0.103 & 1.79 \\ \hline
13 & GT & -0.071 & 0.244 & 0.251 & -0.006 & -0.006 & 2.65 \\ \hline
14 & IBM & -0.113 & 0.598 & -0.359 & 0.310 & -0.114 & 1.72 \\ \hline
15 & IP & -0.208 & 0.244 & 0.3485 & -0.123 & 0.169 & 1.90 \\ \hline
16 & MCD & 0.166 & -0.122 & -0.005 & 0.185 & 0.051 & 2.64 \\ \hline
17 & MRK & 0.244 & -0.288 & 0.039 & 0.243 & 0.149 & 1.93 \\ \hline
18 & MMM & -0.137 & 0.054 & 0.297 & 0.007 & 0.095 & 2.32 \\ \hline
19 & GM & 0.026 & 0.204 & 0.075 & -0.160 & -0.072 & 2.58 \\ \hline
20 & MO & 0.057 & -0.058 & -0.051 & 0.251 & 0.288 & 2.61 \\ \hline
21 & PG & 0.222 & -0.228 & 0.179 & 0.282 & 0.137 & 1.98 \\ \hline
22 & S & 0.079 & 0.018 & 0.060 & 0.009 & -0.002 & 2.86 \\ \hline
23 & TX & -0.599 & -0.328 & -0.266 & -0.062 & -0.073 & 1.12 \\ \hline
24 & UTX & -0.088 & 0.113 & 0.172 & -0.065 & -0.112 & 2.43 \\ \hline
25 & C & 0.174 & -0.036 & 0.054 & -0.145 & -0.395 & 1.60 \\ \hline
26 & HWP & 0.204 & 0.165 & -0.396 & -0.278 & 0.256 & 1.76 \\ \hline
27 & HD & 0.274 & -0.054 & -0.013 & -0.163 & -0.174 & 2.02 \\ \hline
28 & INTC & 0.253 & 0.114 & -0.416 & -0.374 & 0.208 & 1.37 \\ \hline
29 & JPM & 0.153 & -0.046 & 0.106 & -0.128 & -0.405 & 1.71 \\ \hline
30 & JNJ & 0.240 & -0.341 & -0.005 & 0.267 & 0.135 & 1.79 \\ \hline
31 & MSFT & 0.242 & 0.034 & -0.386 & -0.331 & 0.170 & 1.58 \\ \hline
32 & SBC & -0.000 & -0.166 & -0.042 & 0.222 & -0.089 & 2.78 \\ \hline
33 & WMT & 0.231 & -0.172 & 0.028 & -0.076 & -0.164 & 2.16 \\ \hline
34 & DIS & 0.144 & 0.023 & -0.147 & 0.013 & 0.049 & 2.87 \\ \hline
\end{tabular}
\end{center}

\newpage

\begin{figure}[htb]
\begin{center}
\psfig{figure=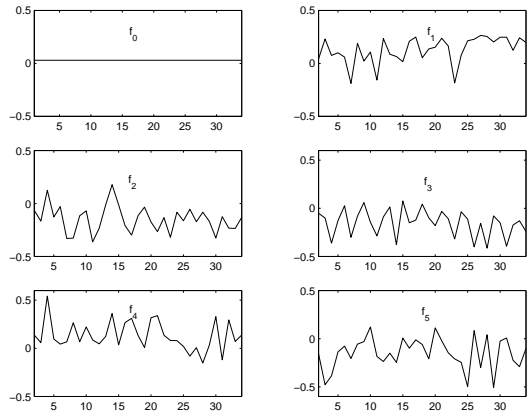,width=7truecm}
\end{center}
\caption{The contribution of each time
series to the factors}
\end{figure}

\section{Clustering and market shocks}

Synchronization in the market plays an important role in the occurrence of
bubbles and crashes. Synchronization it at the root of the disproportionate
impact of public events relative to their intrinsic information content.
This applies to unanticipated public events but also to pre-scheduled news
announcements. Our clustering coefficient, as defined in section 2.3, is
indeed a measure of synchronization in the market and as such may provide
information independent from other market indicators. Not being constructed
from a reduction to a minimum spanning tree, continuous clustering, as we
have defined it, contains maximal information on market
synchronization. As a first step towards a study of the role of this
coefficient we have studied it for a subset of 25 companies, for which we
had much longer time series available. We define {\it volatility} as the
standard deviation of the returns and use centered time windows of 5 and 7
days.

\begin{figure}[htb]
\begin{center}
\psfig{figure=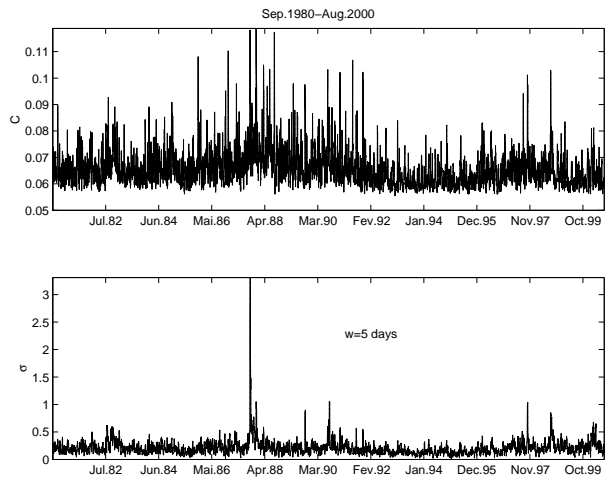,width=8truecm}
\end{center}
\caption{Clustering and volatility for the
period September 1980 - August 2000 with a time window of 5 days}
\end{figure}

In Fig. 6 we compare clustering ($C$) and volatility ($\sigma $) for the
period September 1980 - August 2000 with a time window ($w$) of 5 days. One
notices that most (not all) volatility peaks also correspond to clustering
peaks. However, there are many periods of high clustering which are not
associated to very large volatility. This effect is statistically robust, in
the sense that it remains for much larger time windows. In most cases where
there are simultaneous volatility and clustering peaks, clustering decays
faster than volatility. Although volatility remains high, synchronization
fades out faster after the initial shock. There are exceptions, though (see
below).

\begin{figure}[htb]
\begin{center}
\psfig{figure=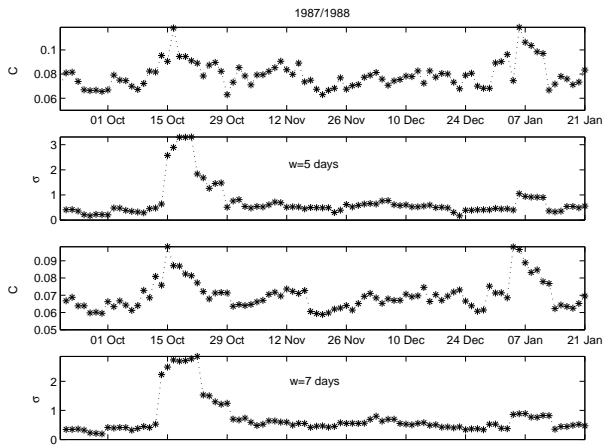,width=8truecm}
\end{center}
\caption{Clustering and volatility for the
period September 1987 - January 1988 with
 time windows of 5 and 7 days}
\end{figure}

\begin{figure}[htb]
\begin{center}
\psfig{figure=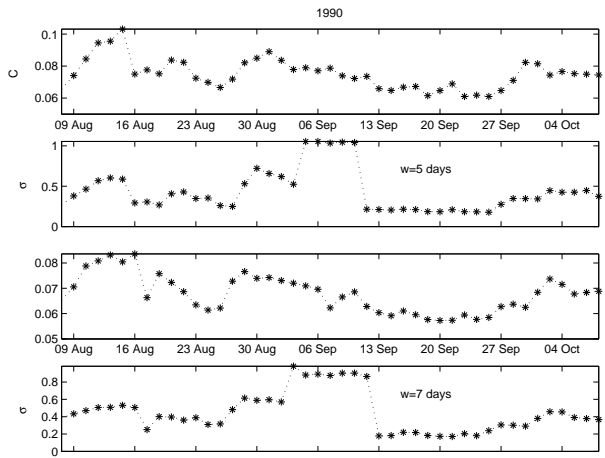,width=8truecm}
\end{center}
\caption{Clustering and volatility for the
period August 1990 - October 1990 with
 time windows of 5 and 7 days}
\end{figure}

\begin{figure}[htb]
\begin{center}
\psfig{figure=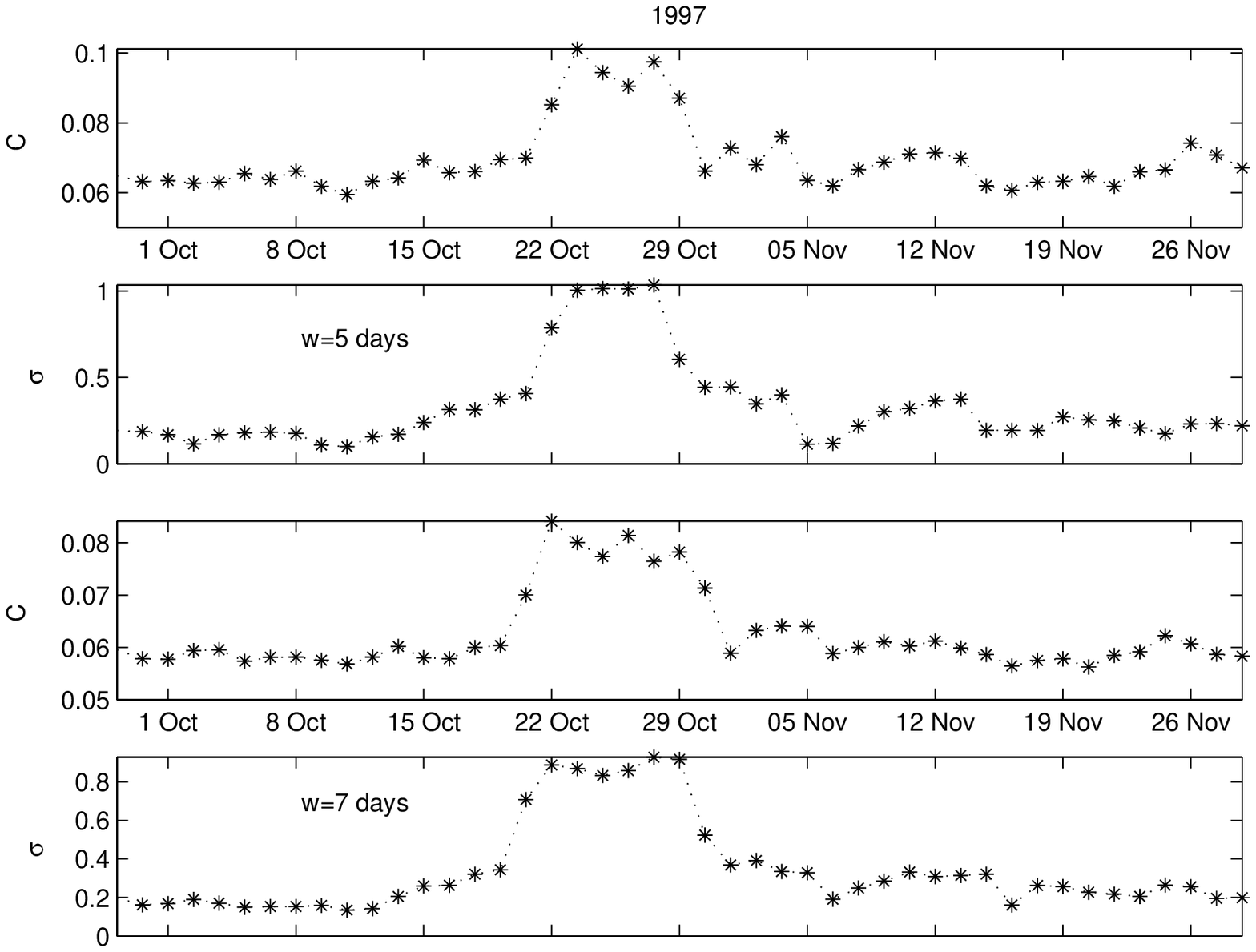,width=8truecm}
\end{center}
\caption{Clustering and volatility for the
period September 1997 - December 1997 with
 time windows of 5 and 7 days}
\end{figure}

In Figs. 7, 8 and 9 we have expanded the periods September 1987 - January
1988, August 1990 - October 1990 and September 1997 - December 1997 using
time windows of 5 and 7 days. In Fig. 7 one sees that around October 19,
1987 (Black Monday) there are both clustering and volatility peaks, but that
clustering (synchronization) decays faster than volatility. In addition
there is around January 6, 1988 another clustering peak that is not
accompanied by exceptionally high volatility. Another interesting example is
provided by Fig. 8 where one sees a clustering peak at around August 15,
1990 with small volatility and a volatility peak after September 5, 1990
without increase in the clustering. Finally, Fig. 9 shows that around
October 27, 1997 (2nd Black Monday - Asian crisis) clustering and volatility
have very similar behavior.

The main conclusion is that clustering indeed provides some new information
on the market which is independent from the one provided by volatility.
Together they provide insight on the different types of market shocks.

\section{Conclusions}

(i) The main result from our empirical study of the market geometric
structure is the dimension reduction that is observed, when compared with
the number of companies of different sectors that are analyzed. This may
have useful implications for economic modelling and the identification of
subspaces and characteristic dimensions may provide a rationale for the
search for {\it economic factors} which are neither sectors nor other
obvious economic facts.

(ii) Underlying all modern views of asset pricing and portfolio selection is
the idea that unsystematic risk may be eliminated by diversification. A
large diversification (comparable to the whole market) involves large costs
and efficient managing. It would be much simpler to have a small number of
partially anticorrelated stocks. In addition to providing a rationale for
the choice of the number of terms in factor models, our approach also
suggests what might be called a dimension-by-dimension (DBD) hedging
strategy, where diversification is not achieved by mimicking the market
portfolio, but by balancing the stocks in appropriate amounts in a few
dimensions.

(iii) In our example (but not necessarily in the method) we have
concentrated on stocks. Nowadays there are on the market a myriad of other
more or less risky assets. In principle the same method also applies to
other financial instruments and it may turn out that the nature of the
economic spaces reconstructed from different asset types will give us
different views on the over-all economic space.

(iv) At a more ambitious level one might think that, once the dimensions of
the economic space are identified, a framework is available to establish
dynamical equations for the market process. However, one should remember that
the bulk of the market fluctuation process seems to be a short-memory
process with a very small long-memory component \cite{Vilela}, which is
nevertheless very important for practical purposes, because it is associated
with the large fluctuations of the returns. Therefore separation of the
components and reconstruction of their characteristic spaces might be an
essential precondition for establishing any meaningful market dynamics
description.

(v) There is a great deal of controversy over experimental tests of the
market efficiency hypothesis in its weak, semistrong and strong versions. At
the theoretical level the modern view of the hypothesis states \cite{Fama}
that market overreaction in some circumstances and underreaction in others
is a pure chance event. In other words, the expected value of abnormal
returns is zero. Other views state that a behavioral component \cite
{Shiller} must always be included in any description of the market.
Behavioral trends, however, may not be inconsistent with a pure statistical
description if the different reaction times and secondary reactions are
taken into account \cite{Olsen}.

Our results do suggest the existence of a certain amount of structure in the
market. However it is a result neither in favor nor against the market
efficiency hypothesis because even if, by careful consideration of the
market structure along the lines we propose in this paper, dimensions and
the ambient manifold become well defined, no conclusion can be drawn on the
nature of the stochastic process that is taking place there.

(vi) Finally, an important spillover from our metric discussion of the
market structure is the notion of continuous clustering which may provide
useful insight on synchronization and market shocks.

{\bf Acknowledgment}

Two of the authors (T. A. and R. V. M.) are grateful for the sponsorship of
the Zentrum f\"{u}r interdisziplin\"{a}re Forschung, Universit\"{a}t
Bielefeld, where part of this research was carried out.

\end{document}